\documentstyle[aps,pre,multicol]{revtex}
\begin{document}
\draft
\date{\today}
\title{Anomalous scaling of triple correlation function of white-advected
passive scalar}
\author{D. Gutman and E. Balkovsky}
\address{
Department of Physics of Complex Systems, Weizmann Inst. of Science,
Rehovot 76100, Israel}
\maketitle
\begin{abstract}
For a short-correlated Gaussian velocity field the problem of a passive 
scalar with the imposed constant gradient is considered. It is shown 
that the scaling of three-point correlation function is anomalous.
In the limit of large dimension of space $d$ anomalous exponent
is calculated.
\end{abstract}

\begin{multicols}{2}

We consider the problem of passive scalar $\Theta$ advected by
delta-correlated in time velocity field ${\bbox v}$ \cite{kr1}. The scalar
obeys an equation
\begin{eqnarray}
\partial_t \Theta + {\bbox v}\nabla \Theta=\kappa \Delta \Theta
\label{i1}\end{eqnarray}
Here $\kappa$ is a molecular diffusivity, which determines behavior of 
passive scalar for small distances. The velocity ${\bbox v}$
is a random function to be defined below. There is no pumping of a
passive scalar into the system, but instead the condition on mean value
of passive scalar is imposed, namely it has a constant gradient in space.
\begin{eqnarray}&&
\Theta={\bbox g}{\bbox r}+\theta.
\label{i2}\end{eqnarray}

Here $\theta$ designate the fluctuating part of a scalar field  
\begin{eqnarray}&& 
\langle \theta \rangle =0.
\label{i3}\end{eqnarray}
By $\langle\rangle$ we mean statistical average.
This problem was previously considered by Shraiman and Siggia in
\cite{sh1}, where they derived equation for the n-point correlation function
and found the pair correlation function. In \cite{sh2,sh3} they
proposed phenomenological equation describing the n-point correlation
functions of passive scalar.

The description of this problem, analogous to Kolmogorov dimensional
analysis gives correct answer for two-point correlation function
$f(r_{12})=\langle\theta(r_1)\theta(r_2)\rangle$.
However it has a deviation from experiment data \cite{mes,ton}
for higher order correlation functions. We will concentrate on such
deviation for the three-point correlation function
$\Gamma=\langle\theta(r_1)\theta(r_2)\theta(r_3)\rangle$, which is the lowest
that has anomalous behavior (i.e. different from one obtained by
dimensional estimate). The imposed gradient of mean value looks like
breaking of translational and rotational invariance. However, the correlation
functions of $\theta$ are translational invariant and the gradient of mean
value of a scalar field will effectively play a role of anisotropic pumping. 
There is a significant difference between this problem and the problem 
with isotropic pumping. Due to the presence of chosen direction in space
the odd order correlation functions do not vanish.

We assume the velocity field to be Gaussian and delta-correlated in time
\cite{kr1}. Therefore statistics of ${\bbox v}$ is completely determined
by the pair-correlation function
\begin{eqnarray}&& 
\left\langle v^{\alpha}(r_1,t_1)v^{\beta}(r_2,t_2)\right\rangle=
V^{\alpha \beta}(r_{12}) \delta(t_1-t_2)\\&&
V^{\alpha \beta}(r)=V_0 \delta^{\alpha \beta}-K^{\alpha \beta}(r) 
\label{i4}\end{eqnarray}
Here $K^{\alpha\beta}$ is the eddy diffusivity tensor 
\begin{eqnarray}&&
K^{\alpha \beta}(r)=
Dr^{-\gamma}
\left[\frac{d+1-\gamma}{2-\gamma}r^2\delta^{\alpha\beta}-r^{\alpha}r^{\beta}
\right],
\label{i5}\end{eqnarray}
where $d$ is the dimension of space and $\gamma$ is parameter which
determines the scaling of velocity pair-correlation function.
The parameter $\gamma$ is supposed to be  between zero and two. 
Power-like behavior (\ref{i5}) is valid up to some $r\sim L$, where
$L$ is the correlation length of velocity. For $r>L$ the correlator
$V^{\alpha \beta}$ decreases and goes to zero as $r\to\infty$. 
As we will see later the precise form of this decrease is
not important. Therefore we can choose arbitrary form of pumping
at $r>L$. We will believe $K^{\alpha \beta}(r)=V_0 \delta^{\alpha \beta}$
for large $r$. The relation between constants is
\begin{eqnarray}
L^{2-\gamma}=\frac{2-\gamma}{D(d+1-\gamma)}V_0
\label{i15}\end{eqnarray} 
We assume that the cut-off length $ L $ is d-independent, so that
$V_0$ explicitly depends on space dimensionality.

Our aim is to find the scaling exponent of the
triple correlation function deep inside convective interval, that is for
$r_d\ll r\ll L$, where $r_d$ is diffusion scale:
\begin{eqnarray}&&
r_d^{2-\gamma}=\frac{2\kappa(2-\gamma)}{D(d-1)}.
\label{i13}\end{eqnarray} 
The condition $r\gg r_d$ implies that we can disregard diffusion \cite{CFKL95}.
 
From (\ref{i1}) one can derive the equation for the nth-order correlation 
function \cite{kr1,sh1}
\begin{eqnarray}
\hat{\cal L} \langle \Theta_1...\Theta_n \rangle=0.
\label{i6}\end{eqnarray}
Here the operator $\hat{\cal L}$ contains both operators of turbulent and
molecular diffusion:
\begin{eqnarray}
\hat{\cal L}=\frac{1}{2}\sum_{i,j} V^{\alpha\beta}(r_{ij}) \nabla^\alpha_i 
\nabla^\beta_j+\kappa\sum_i \nabla_i^2
\label{i7}\end{eqnarray}

We will see that the scaling of the pair-correlation function is normal
that is $\gamma$. For the three-point correlation function the presence
of zero modes of the operator $\hat{\cal L}$ makes scaling anomalous
that is different from $1+\gamma$  obtained from Kolmogorov-like estimates.
The exception is special case when the scaling exponent of the velocity
correlation function $\gamma$ is zero \cite{sh1}. Note also that for the
scaling exponent equal to two the operator $\hat{\cal L}$ has a singularity
and special treatment is needed \cite{pum}. Following \cite{CFKL95} we
will calculate anomalous scaling exponent in the limit of large $d$.
For $d\to\infty$ the problem can be solved exactly. Then, finding
corrections in the next order over $1/d$ one can calculate the anomalous
exponent of the three-point correlator.

To calculate the triple correlation function one should know the two-point
correlator since it explicitly enters the equation. 
The equation for the two-point correlation function is 
\begin{eqnarray}
\hat{\cal L}^{(p)}f(\bbox{r}_{12})=-\hat{\cal L}\left [ ({\bbox g}{\bbox r}_1)
({\bbox g}{\bbox r}_2) \right ]
\label{p1}\end{eqnarray}
The operator $\hat{\cal L}^{(p)}$ may be written as 
\begin{eqnarray}
\hat{\cal L}^{(p)}=
\frac{D(d-1)}{2-\gamma}
r^{1-d}\partial_r \left(r^{d+1-\gamma}\partial_r\right), \, \ r_d\ll r<L 
\label{p3a}\end{eqnarray}
We should match the solution obtained at $r<L$ with the solution at $r>L$.
For this region the operator has the following form
\begin{eqnarray}&&
\hat{\cal L}^{(p)}=V_0 r^{1-d}\partial_r \left(r^{d-1}\partial_r\right)
\label{p3}\end{eqnarray}
Since we will consider three-point correlation function at large $d$,
we calculate pair-correlation function in the same limit
(it can be found exactly as well).
The solution of equation (\ref{p1}) at $r<L$ and $d\gg1$ which satisfies
boundary condition at $r\sim L$ is
\begin{eqnarray}&&
f(r)=\frac{g^2L^2}{d}\left[
\frac{2-\gamma}{2\gamma}-
\frac{1}{\gamma}\left(\frac{r}{L}\right)^{\gamma}+
\frac{1}{2}\left(\frac{r}{L}\right)^2
\right]
\label{p8}\end{eqnarray}
At $r>L$ the pair correlation function is zero
in the main order of $1/d$ expansion.
For $r$ deep inside the convective interval $r^2$ term is small in comparison 
with $r^\gamma$. Therefore we see that the
scaling behavior of a two-point correlation function is normal, as it was
shown by Shraiman and Siggia \cite{sh2}.
Note that (\ref{p8}) is isotropic in convective interval. 
The anisotropic part has scaling exponent larger then $\gamma$ and therefore
in the convective interval is much smaller. Besides, it has additional factor
$1/d$ with respect to isotropic part.

For the three-point correlation function one can derive the
following equation
\begin{eqnarray}&&
\hat{\cal L}\Gamma=\Phi_{12,3}+\Phi_{13,2}+\Phi_{23,1},
\,\mbox{where}\label{t1}\\&&
\Phi_{12,3}=\nabla_1^{\alpha}f(r_{12})g^{\beta}
\left[
K^{\alpha\beta}(r_{13})-K^{\alpha\beta}(r_{23})
\right].
\label{t1a}\end{eqnarray}

Since the preferable direction exists in the problem then $\Gamma$ depends on
five independent variables in any dimension of space $d>2$.
It is convenient to choose the following set
\begin{eqnarray}&&
\Gamma=\Gamma\left(r_{12},r_{13},r_{23},{\bbox r}_{12}{\bbox g},
{\bbox r}_{23}{\bbox g}\right).
\label{t4}\end{eqnarray}
However, this choice is is not unique because of identity
${\bbox r}_{12}+{\bbox r}_{13}+{\bbox r}_{23}=0$. 

Unfortunately, the number of variables is too large to solve the 
equation (\ref{t1}) exactly at arbitrary $\gamma$ and $d$.
There are several cases when the problem can be solved exactly:
the cases of $\gamma=0$ \cite{CFKL95a,BCKL95,sh2}, of large $d$
\cite{CFKL95}, and of $\gamma=2$ \cite{GK}. After solving
the problem exactly at these special values of parameters one can develop
the perturbation theory for the small deviations from these values.
In the cases of large $d$ and $\gamma=2$ this perturbation theory can be
performed. The case of $\gamma=0$ corresponds to the strongly
degenerate perturbation theory.
We are going to solve the equation (\ref{t1}) in the limit of large $d$,
where perturbation theory is regular.
To do this one should keep in the operator (\ref{i7}) only
terms which are the largest at large $d$ and solve the equation.
Then one can look for a correction due to a finite value of the parameter
$1/d$. Justification of such a perturbation theory can be found in
\cite{CFKL95}.

In variables (\ref{t4}) the main part of the operator $\hat{\cal L}$
can be written as
\begin{eqnarray} 
\hat{\cal L}_0=\frac{d^2 D}{2-\gamma}
\sum_{i<j}r_{ij}^{1-\gamma}\partial_{r_{ij}}.
\label{t3}\end{eqnarray}
For large $d$ and $r\ll L$ one can substitute (\ref{p8}) into the 
right-hand side of (\ref{t1}). The result is
\begin{eqnarray}&&
\Phi_{12,3}^{(0)}=-\frac{g^2DL^{2-\gamma}}{2-\gamma}\left[
({\bbox r}_{12}{\bbox g})r_{12}^{\gamma-2}(r_{13}^{2-\gamma}-
r_{23}^{2-\gamma})
\right]
\label{t2}\end{eqnarray}
Thus, on the first step of iteration procedure we should solve the
following equation
\begin{eqnarray}&&
\hat{\cal L}_0\Gamma_0=\Phi_{12,3}^{(0)}+\Phi_{13,2}^{(0)}+\Phi_{23,1}^{(0)}
\label{t5}\end{eqnarray}
The solution can be found by integrating over the characteristics of
the operator (\ref{t3}). Important question is that of boundary
conditions. We should supply equation (\ref{t5}) with the boundary conditions
at zero and infinity. However this equation is of the first order and only
one condition can be satisfied. Since we omitted diffusion part of the
operator (\ref{i7}) we should pose the only condition
$\Gamma\to 0$ as $r\to\infty$. One can show \cite{CFKL95} that solution 
obtained in this way can be matched with diffusion region.

One can write the following solution of (\ref{t5})
\begin{eqnarray}&& 
\Gamma_0=\frac{g^2L^{2-\gamma}}{\gamma d^2}({\bbox r}_{12}{\bbox g})
\int_0^{L^{\gamma}}\!\!
dt\,(r_{12}^\gamma+t)^{1-2/\gamma}\nonumber\\&& 
\times\!\left[(t+r_{13}^\gamma)^{2/\gamma-1}-
(t+r_{23}^\gamma)^{2/\gamma-1} \right]+\mbox{permutations}
\label{t6}\end{eqnarray}
This integral diverges on upper limit and we have to
regularize it. Since divergence is logarithmic then the precise form of
regularization is not important, we just cut off the integral at
$t\sim L^{\gamma}$. It corresponds to the condition that correlator
is zero for $r>L$. Integral (\ref{t6}) can not be calculated
analytically for arbitrary $\gamma$. However we are interested
in $\Gamma$ in the inertial interval, that is for $r\ll L$. There
main contribution to (\ref{t6}) comes from large $t$. Expanding
integrand over $r/L$ we find the following result in the main logarithmic
order
\begin{eqnarray}&& \label{a20}
\Gamma_{0}=\frac{2(2-\gamma)}{d}\ln{\left(\frac{L}{r}\right)}Z_0,
\label{t8}\end{eqnarray}
where $r$ in the logarithm is of order $r_{ij}$ and 
\begin{eqnarray}&& 
Z_0=\frac{g^2 L^{2-\gamma}}{2\gamma d} 
\Bigl[({\bbox r_{12}}{\bbox g})(r_{13}^{\gamma}-r_{23}^{\gamma})\nonumber\\&&+
({\bbox r_{13}}{\bbox g})(r_{12}^{\gamma}-r_{23}^{\gamma})  
+({\bbox r_{23}}{\bbox g})(r_{12}^{\gamma}-r_{13}^{\gamma})\Bigr] 
\label{t7}\end{eqnarray}
Note that $Z_0$ is a zero mode of the operator (\ref{t3}). It is necessary
since logarithm enters the solution. If it were not zero mode, we would have
logarithm at the right-hand side of (\ref{t1a}). It is also worth mentioning
that $Z_0$ has scaling $\gamma+1$ and this is the only zero mode of the
operator $\hat{\cal L}_0$ with such a scaling.

Now we should find a correction to (\ref{t8}) in the next order over $1/d$.
To do this we should solve the equation
\begin{eqnarray}&&
\hat{\cal L}_0 \Gamma_1=-\hat{\cal L}_1\Gamma_0.
\label{t9}\end{eqnarray}
Here $\hat{\cal L}_1$ is the part of operator (\ref{i7}) which is proportional
to $d$. In equation (\ref{t9}) we disregard terms which come from the
right-hand side of the equation (\ref{t1}) since they do not contain
logarithm. Solving equation (\ref{t9}) we find for the main logarithmic term
\begin{eqnarray}&&
\Gamma_1=\frac{2(2-\gamma)^2}{d^2}\,Z_0\,\ln^2\left(\frac{L}{r}\right).
\label{t11}\end{eqnarray}
We can continue this iteration procedure and find the main 
logarithmic subsequence. As a result we obtain the following series
\begin{eqnarray}&&
\Gamma=Z_0\left(1+\Delta\ln{\left(\frac{L}{r}\right)}+
\frac{\Delta^2}{2}\ln^2{\left(\frac{L}{r}\right)}+\ldots\right)-Z_0
\nonumber\\&&
=Z_0\left(\frac{L}{r}\right)^{\Delta}-Z_0,\label{t12}
\end{eqnarray}
where the anomalous exponent $\Delta$ which makes deviation from
normal scaling is 
\begin{eqnarray}&&
\Delta=\frac{2(2-\gamma)}{d}.
\label{t14}\end{eqnarray}
We see, that the solution consists of two parts. The second term
in the right-hand side of (\ref{t12}) has normal scaling $\gamma+1$.
One can check that it is
partial solution of inhomogeneous equation (\ref{t1}).
However, if it were alone, the solution would not satisfy boundary conditions.
To ensure it we have the first term in (\ref{t12}). One can see that
this contribution gives anomalous scaling with exponent $\gamma+1-\Delta$.
This term is the solution of homogeneous equation $\hat{\cal L} \Gamma=0$
and therefore is the zero mode of the operator $\hat{\cal L}$
\cite{CFKL95,GK,sh2}.

To conclude, we have shown that the three-point correlation function has
the scaling exponent $\gamma+1-\Delta$ which differs from naive
dimensional estimates. This exponent was analytically calculated in the
leading $1/d$ order.

\acknowledgements
We are grateful to M. Chertkov, G. Falkovich, I. Kolokolov, and
V. Lebedev for fruitful discussions.

\end{multicols}

\end{document}